\documentclass{aa501}
\usepackage{graphics}
\usepackage{graphicx}
\usepackage{amssymb}
\begin{document}

\def\ls{\,^<\!\!\!\!_\sim\,\,}
\def\gs{\,^>\!\!\!\!_\sim\,\,}
\def\pn{\par\noindent}
\def\ss{\smallskip\pn}
\def\ms{\medskip\pn}
\def\bs{\bigskip\pn}
\def\ai{$\alpha_{\rm inj}$ } 	    %alpha injection
\def\vb{$\nu_{\rm br}$ }	    %break frequency

\title{A multi-Frequency Study of The Radio Galaxy NGC326.}
\subtitle{I. The Data}

\author{M. Murgia\inst{1,2}
\and P. Parma\inst{1}
\and H. R. de Ruiter \inst{1,4}
\and M. Bondi \inst{1} 
\and  R. D. Ekers \inst{5}
\and R. Fanti \inst{1,3}
\and E. B. Fomalont \inst{6}
}

\institute{Istituto di Radioastronomia del CNR, Via Gobetti 101, I-40129, 
Bologna, Italy
\and Dipartimento di Astronomia, Universit\`a di Bologna, Via Ranzani 1,
I-40127 Bologna, Italy
\and Dipartimento di Fisica, Universit\`a di Bologna, Via B. Pichat 6/2,
 I--40127 Bologna, Italy
\and Osservatorio Astronomico di Bologna, Via Ranzani 1,I-40127 Bologna, Italy
\and Australia Telescope National Facility, CSIRO, P.O. Box 76, Epping, NSW 
2121, Australia
\and National Radio Astronomy Observatory, 520 Edgemont Road, Charlottesville,
 VA 2293
}

\date{Received; Accepted}

\abstract{
We present the results of a multi-frequency study of the inversion 
symmetric radio galaxy NGC326 based on Very Large Array  observations
 at 1.4, 1.6, 4.8, 8.5 and 14.9 GHz. The morphological, spectral and
 polarization  properties of this peculiar object are studied  at different
 levels of spatial resolutions. The interpretation of the data will be
 discussed in forthcoming papers.
}

\offprints{M. Murgia, \email{murgia@ira.bo.cnr.it}}

\titlerunning{Multi-Frequency Study of The Radio Galaxy NGC326. I. The Data}
\authorrunning{M. Murgia et al.}
\maketitle

\section{Introduction}
The radio source B2 0055+26 was identified with the elliptical galaxy NGC326
 during a program directed at the identifications of the optical 
counterparts of radio sources selected from the B2 catalogue (\cite{Colla75}).
Basic properties of NGC326 are reported in Table ~1.
The radio source was first mapped by Fanti et al. (1977) at 1.4 GHz with
 the Westerbork Synthesis Radio Telescope (WRST) in the snapshot mode. 
Because of the interesting structure seen, this observation was followed up by 
a more sensitive WRST observation at 5 GHz with an angular resolution of
 $6\arcsec \times 13\arcsec$ ~(\cite{Ekers78}). At this resolution the radio
 source was found to be composed of two curving tails with a striking 
180\degr \, symmetry. NGC326 was recognized as one of the most spectacular 
examples of inversion symmetric radio galaxy. The authors explained the unusual
 structure as being due to the slow precession of the radio jets during the lifetime
 of the radio source.
Rees (1978) suggested that the beam direction  precesses
 due to a realignment caused by the accretion of gas with a different
 angular momentum direction with respect to the central black hole axis.\\

Distorted radio galaxies can by classified according to two broad schemes: 
the mirror symmetric (`C'-shape) and the inversion symmetric (`Z'-shape).
 A sub-class, called `X'-shape, is characterized by two separate low
 brightness wings, almost perpendicular to the currently active lobes.
 There is a broad consensus concerning the dynamical interpretation of
 the `C'-shape.
 This kind of distortion is either caused by the translational motion of the
 galaxy through the intergalactic medium (wide and narrow angle tails) or
 by the orbital motion of the galaxy around a nearby companion
 (\cite{BlandfordIcke78}). The explanation of the `Z'-shape and `X'-shape 
is still uncertain. Along with the secular
 jet precession (\cite{Rees78}), other two alternative scenarios
 have been proposed: the sudden realignment of the jet (Wirth et al. 1982)
 and the buoyancy of the material from the lobes (\cite{Worrall95}).\\
Battistini et al. (1980) found that NGC326 is 
a double system composed of two nearly equally bright elliptical galaxies in
a common envelope (``dumbbell'' galaxy).
 \cite{Wirth82} investigated the connection between the optical and
 the radio morphology for a sample of $\sim$100 radio emitting dumbbell 
galaxies. They found about a dozen
 of sources with `Z' or `X'-shape and suggested that the presence of another
 equally massive galaxy within 10-30 kpc of a radio galaxy would strongly 
influence the jet properties.
 Particularly, they argue that in the case of bound circular orbits the
 continuous tidal interaction can produce mirror-symmetric
 wiggling jets as previously  suggested by Blandford \& Icke (1978).
 In the case of unbound orbits an impulsive interaction between the two
 galaxies can lead to an inversion-symmetric radio source like NGC326. 
The 'Z'-shape occurs when the duration of the impulsive torque is longer 
than the jet outflow time to the lobes, while the 'X'-shape results when
 the crossing time is so fast, as compared to the jet outflow time, as to 
cause an instantaneous change in the jet direction.
 \cite{Wirth82} studied in detail the case of NGC326,
 using a new optical image.
By examining the isophotes they found  a brightness difference $<$1 mag
 between the two galaxies and very little deviation from pure ellipticity.
 Indeed they proposed NGC326 as a prototype for the class of
 undisturbed dumbbell galaxies. They also pointed the
 attention to  the quite high velocity difference between the two galaxies:
 $\Delta v \simeq 750$ km\,s$^{-1}$ (Sargent 1973).
 In the meantime the first 20-cm Very Large Array (VLA)
 image of the radio source became available (\cite{Fomalont81},
 \cite{Ekers82}). In the light of the new optical and radio data,
 \cite{Wirth82} reviewed the morphological classification and the
 interpretation of the source made by Ekers et al. (1978). They suggested
 that NGC326 describe an `X'-shape in which the old jet direction
was from NE to SW and the actual jet direction is from SE to NW. Moreover,
both the old and the new lobes show a `C'-symmetry but with a difference in
the position angle of about 120\degr \,, suggesting that the radio galaxy
has abruptly changed its velocity direction during its transient interaction
with the other galaxy. In the Wirth et al. (1982) picture the velocity of 
the jet flow cannot be much grater than the $\Delta v$ of the galaxies. More
 recent redshift measures give $\Delta v=509\pm 44 ~\rm km ~\rm s^{-1}$
 (\cite{Davoust95}) and  $\Delta v=549\pm 38 ~\rm km ~\rm s^{-1}$
 (\cite{Werner99}).\\

NGC326 is the brightest member of a small group of galaxies, 
Zwicky 0056.9+2636. Werner et al. (1999) measured the redshift for eight
 galaxies of the group (including NGC326) confirming the presence of a 
cluster with a mean redshift of $z_{\rm mean}=0.0477 \pm 0.0007$ and 
line-of-sight velocity dispersion $\sigma_{z}=599^{+230}_{-110}$ km\,s$^{-1}$.
Werner et al. (1999) showed that the brightest of the two optical galaxies 
(the radio source) is a slowly-moving member of the cluster while its
 companion has a velocity of about 500 km~s$^{-1}$ relative to the 
cluster velocity.

 The first detailed studies of the intergalactic medium
 surrounding NGC326 were done by Worrall et al. (1995) and 
Worrall \& Birkinshaw (2000). The region containing the source
was imaged in soft X-ray with the \emph{ROSAT} PSPC. Surprisingly, they 
discovered that the galaxy is embedded in a bright asymmetric
 X-ray-emitting gaseous medium with properties more typical of a cluster, 
rather than of a group. They report a gas temperature of kT $\sim 2$
keV and a 0.1-2.4 keV luminosity of $7.5 \times 10^{36} ~h^{-2}$ W. From
 the PSPC image the emitting gas has a full extent of $\ge 400 ~h^{-1}$ kpc,
 stretching north-east of the peak, which  coincides with the position
of NGC326. The best description of the radial profile they obtained is a 
combination of a cluster $\beta$-model and an unresolved component
 (possibly a hot galactic corona). They obtained a cluster core radius of
 171 $h^{-1}$ kpc. Because of the presence of the compact hot corona
 centered on NGC326, Worrall et al. (1995) proposed that buoyancy effects
 of the radio plasma in the X-ray atmosphere could be responsible for the
 bending of the radio lobes. They suggested that, in projection, each radio
 lobe bends sideways into a tail as the backflow approaches the galactic
 corona. In order to explain the observed source shape, the buoyancy condition
 requires a backflow velocity for the lobes material of $v \simeq 3000$
 km\,s$^{-1}$.\\

\begin{table}[t]
\label{thesource}
\caption[]{Basic properties of NGC326.}
\begin{tabular}{ll}
\hline
\noalign{\smallskip}
Optical position $(\alpha_{\rm J2000})$&  00$^{h}$58$^{m}$22\farcs6 \\
Optical position $(\delta_{\rm J2000})$& +26\degr51\arcmin58\farcs3 \\
Redshift           			&0.0477\\
Distance				&141~$h^{-1}$ Mpc\\ 
Magnitude (B$_{{\rm T}^{0}}$)              &13.9\\
1.4~GHz total flux density	& 1.77 Jy\\
1.4~GHz total radio luminosity		& $10^{24.6}h^{-2}$ W/Hz\\
Integrated rotation measure             & $-25\pm 5$ rad ~m$^{-2}$\\
Overall spectral index                  & 0.8\\
Radio source largest linear size        & 180~$h^{-1}$ kpc\\ 
Arcsec to kpc conversion factor& 1\arcsec = 0.63 $h^{-1}$ kpc\\
\noalign{\smallskip}
\hline
\multicolumn{2}{l}{\scriptsize $h={\rm H}_{0}/100_{\rm ~km~s^{-1} 
Mpc^{-1}}$, $q_{0}=0.5$}\\
\multicolumn{2}{l}{\scriptsize We use the convention $S_{\nu}\propto 
\nu^{-\alpha}$}\\
\end{tabular}
\end{table}
This is paper I of a series of papers dedicated to the radio galaxy NGC326.
Here we give  a description of the morphological, spectral
 and polarization characteristics of this radio galaxy.
In  Sec. 2 we summarize the observations used in this paper.
A description of the source morphology and its connections with the
 polarization and spectral properties is presented in Sec. 3.
 In Sec. 4 and 5 we analyzed the jet collimation properties and the source
 physical parameters, respectively. A summary of the paper is given
 in Sec. 6.\\  

 We will discuss the interpretation of the spectral and polarization 
properties respectively in the forthcoming papers II and III.

\section{The VLA observations and data reduction}
A summary of the observations, including the VLA configuration,
frequency, bandwidth, date and length of observations, is reported in Table ~2.

\begin{table}[h]
\caption[]{VLA observations summary.}
\label{VLAobs}
\begin{center}
\begin{tabular}{ccccc}
\hline
\noalign{\smallskip}
   Array & Frequency & Bandwidth & Date &  Duration   \\
         & GHz     & MHz     &  &          hours     \\
\noalign{\smallskip}
\hline
\noalign{\smallskip}
  A &  1.41/1.66 & 12.5& 02-Dec-1984 & 8.0 \\
  A &  4.86       & 12.5& 03-Dec-1984 & 1.0 \\
  B &  4.86       & 25.0& 22-May-1985 & 3.0 \\
  B &  14.96     & 25.0& 22-May-1985 & 0.9 \\
  C &  1.41/1.66 & 50.0& 28-Sep-1985 & 0.5 \\
  C &  4.86       & 50.0& 28-Sep-1985 & 0.5 \\
  C &  14.96     & 50.0& 28-Sep-1985 & 7.1 \\
  D &  4.83       & 50.0& 28-Sep-2000 & 0.5 \\
  D &  8.46       & 50.0& 28-Sep-2000 & 1.4 \\
  D &  14.96     & 50.0& 23-Nov-1985 & 0.4 \\
\noalign{\smallskip}
\hline
\end{tabular}
\end{center}
\end{table}
The flux densities were brought to the scale of Baars et al. (1997) using 3C48
as primary flux density calibrator. The instrumental polarization and the
 polarization position angle were calibrated using the
 sources 0116+319 (0042+233 for 6 and 3.5~cm D array) and 3C138 respectively.
 The 3C138 polarization angle was assumed to have a value of -18\degr \,
 at L band and -24\degr \,at higher frequencies.
Post calibration reduction was done using the National Radio Observatory
 (NRAO) AIPS package. The data for each observation were reduced using the 
standard self-calibration and cleaning procedures.
The C array observation at 20~cm, already presented by Parma et al. (1991),
 has been reanalyzed for the purposes of this project.

\begin{table}[h]
\caption[]{Image parameters summary.}
\label{imagesSum}
\begin{center}
\begin{tabular}{cccc}
\hline
\noalign{\smallskip}
   Array & Frequency & Beam      & $\sigma_{\rm I}$   \\
         & GHz       & arcsec    & mJy/beam  \\
\noalign{\smallskip}
\hline
\noalign{\smallskip}
  A &  1.41    & $1\farcs47 \times 1\farcs 24$ & 0.05   \\
  A &  1.66    & $1\farcs32 \times 1\farcs05$ & 0.05     \\
  B &  4.86     & $1\farcs19 \times 1\farcs14$ & 0.029  \\
  C &  1.41    & $13\farcs82 \times 12\farcs77$ & 0.024  \\
  C &  1.66    & $11\farcs51 \times 10\farcs71$ & 0.021  \\
  C &  4.86     & $3\farcs99 \times 3\farcs77$ & 0.025  \\
  C &  14.96   & $1\farcs46 \times 1\farcs25$ & 0.034  \\
  D &  4.83     & $14\farcs08 \times 12\farcs81$ & 0.06  \\
  D &  8.46     & $8\farcs97 \times 7\farcs83$ & 0.05  \\
  D &  14.96   & $7\farcs31 \times 4\farcs16$ & 0.08  \\
\noalign{\smallskip}
\hline
\multicolumn{4}{l}{\scriptsize Col. 4: total intensity image noise.}
%\\
\end{tabular}
\end{center}
\end{table}

The angular size of NGC326 exceeds 4.4 arcmin. This represents a problem
for high frequency imaging since the radio structure size is comparable to
the field of view of a single VLA antenna. The old 2~cm observations
suffered from this effect. Learning from experience, we followed a
 specific observation strategy with the more recent 3.5~cm observations.
 In order to reduce the problems due to the strong primary beam attenuation 
we splitted the observations into two different pointings centered at
 appropriate positions on the wings of the source. The two data sets were
 reduced independently and the resulting couple of images was then combined
 using the task {\it LTESS} in AIPS. Table ~3 summarize the relevant
 parameters of the single-array images. 

 Data from different arrays were combined to improve uv-coverage and 
sensitivity. We combined the A and C arrays at 20~cm,
the B, C and D at 6~cm, and the C and D at 2~cm. Each combined
 data set was self-calibrated. The images at different frequencies 
were cleaned and restored with the same beam using the AIPS 
task {\it IMAGR}. The ``original'' beams were very similar
 to the finally adopted ones. For the purposes of the spectral and
 polarization analysis we  made data sets composed of matched resolution
 images, using natural and uniform weighting. Finally we obtained four
 distinct resolution data sets with exactly the same
cut in the maximum uv baseline at each frequency.
 Their resolutions are 10\arcsec, 4\arcsec, 2\arcsec, and 1\farcs2. 

\begin{table}[h]
\label{mapcube}
\caption[]{Combined array image parameters. The uv weighting function is
 indicated just after the frequency. UW and NW stand for uniform and
 natural weights, respectively.}
\begin{center}
\begin{tabular}{cccc}
\hline
\noalign{\smallskip}
   Array & Frequency     & $\sigma_{\rm I}$ &  $\sigma_{\rm Q,U}$   \\
         & GHz          & mJy/beam         &  mJy/beam   \\
\noalign{\smallskip}
\hline
\scriptsize{$10\arcsec \times 10\arcsec$ beam}&&&\\
\noalign{\smallskip}
  A+C &  1.41  {\tiny UW}&   0.22  & 0.13 \\
  A+C &  1.66  {\tiny UW}&   0.24  & 0.14 \\
  C+D &  4.86  {\tiny NW}&   0.07  & 0.06 \\
    D &  8.46  {\tiny NW}&   0.07  & 0.01 \\
  C+D &  14.96 {\tiny NW}&   0.20  & 0.07 \\
\noalign{\smallskip}
\hline
\scriptsize{$4\arcsec \times 4\arcsec$ beam}&&&\\
\noalign{\smallskip}
  A+C &  1.41 {\tiny NW}   &  0.07    & 0.04 \\
  A+C &  1.66 {\tiny NW}   &  0.08    & 0.05 \\
  B+C &  4.86 {\tiny NW}   &  0.05    & 0.03 \\
  C+D &  14.96 {\tiny NW}  &  0.04    & 0.04 \\
\noalign{\smallskip}
\hline
\scriptsize{$2\arcsec \times 2\arcsec$ beam}&&&\\
\noalign{\smallskip}
  A+C &  1.41 {\tiny NW}    & 0.04   & 0.03 \\
  A+C &  1.66 {\tiny NW}    & 0.05   & 0.04 \\
  B+C &  4.86 {\tiny NW}    & 0.04   & 0.03 \\
  C+D &  14.96 {\tiny NW}   & 0.03   & 0.03 \\
\noalign{\smallskip}
\hline
\scriptsize{$1\farcs2 \times 1\farcs2$ beam}&&&\\
\noalign{\smallskip}
  A &  1.41 {\tiny UW}    & 0.06   & 0.06 \\
  A &  1.66 {\tiny UW}    & 0.07   & 0.07 \\
  B+C &  4.86 {\tiny UW}  & 0.04 & 0.05 \\
  C+D &  14.96 {\tiny UW} & 0.04 & 0.06 \\
\noalign{\smallskip}
\hline
\multicolumn{4}{l}{\scriptsize Col. 3: total intensity image noise;}\\
\multicolumn{4}{l}{\scriptsize Col. 4: polarization image noise.}
\end{tabular}
\end{center}
\end{table}

\begin{figure*}
\begin{center}
\includegraphics[angle=-90, width=18cm]{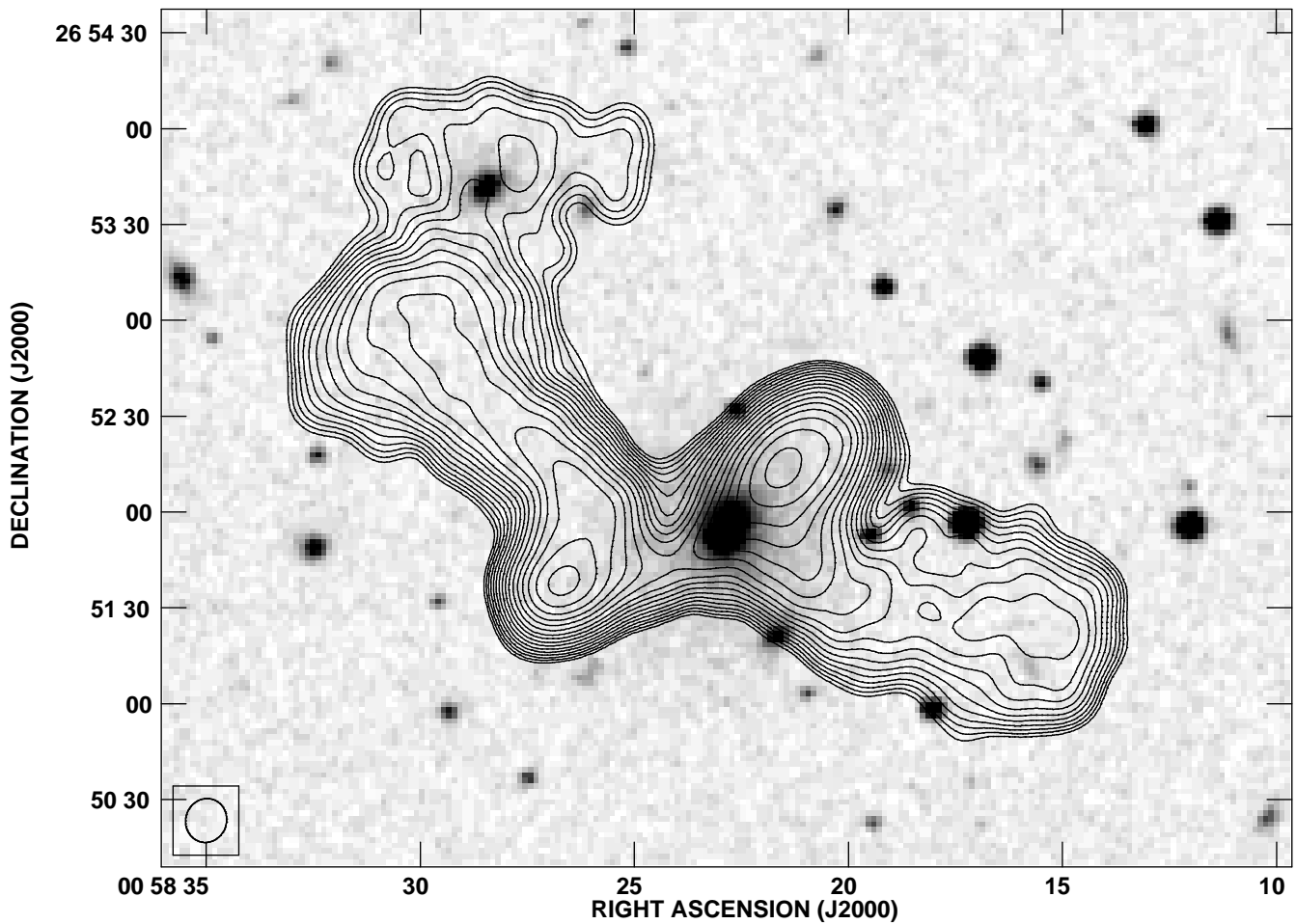}
\end{center}
\caption[]{1.4 GHz C array (contours) overlayed onto the optical 
image from the red Palomar Digitized Sky Survey. The contour levels are 0.5,
 0.71, 1, 1.4, 2, 2.8, 4, 5.7, 8, 11, 16, 23, 32, 45, 64, 91, 130, 180 mJy    
 (beam area)$^{-1}$ and the restoring beam is $13\farcs8 \times 12\farcs 8$ 
with position angle $-16.9\degr$.}
\label{C.14}
\end{figure*}

The equal resolution data sets are given in Table ~4.
All the images were corrected for  primary beam attenuation.
NGC326 is significantly polarized at all frequencies and resolutions.
We produced images of the scalar fractional polarization ($P_{\nu}$),
 depolarization ($DP^{\nu_{1}}_{\nu_{2}}$, defined as the ratio between 
the fractional polarization at $\nu_{1}$ and $\nu_{2}$) and rotation
 measure (RM). The polarization images were corrected for the non-Gaussian
 noise distribution of the polarized intensity.
The RM was obtained using the AIPS task {\it RM} by a weighted fit of the 
position angle to the square wavelength at four frequencies. 

The spectral analysis has been performed with the program Synage++
 (Murgia 2001).  

\section{Source morphology, spectrum and polarization}
The observations described in the previous section allow us to investigate the
morphology of NGC326 at increasing levels of angular resolution. 
In this section we describe the source morphology and its connections with 
the polarization and spectral properties going from the extended structures 
(low resolution images) to the finest details (high resolution images).
 We refer to the different regions of the radio source as shown in Fig.~3.
 In particular, the two ellipses indicate the separation we marked between 
lobes and wings.

\begin{figure*}[t]
%BoundingBox: 9 100 605 560
%was 10cm
\includegraphics[width=10cm, angle=-90]{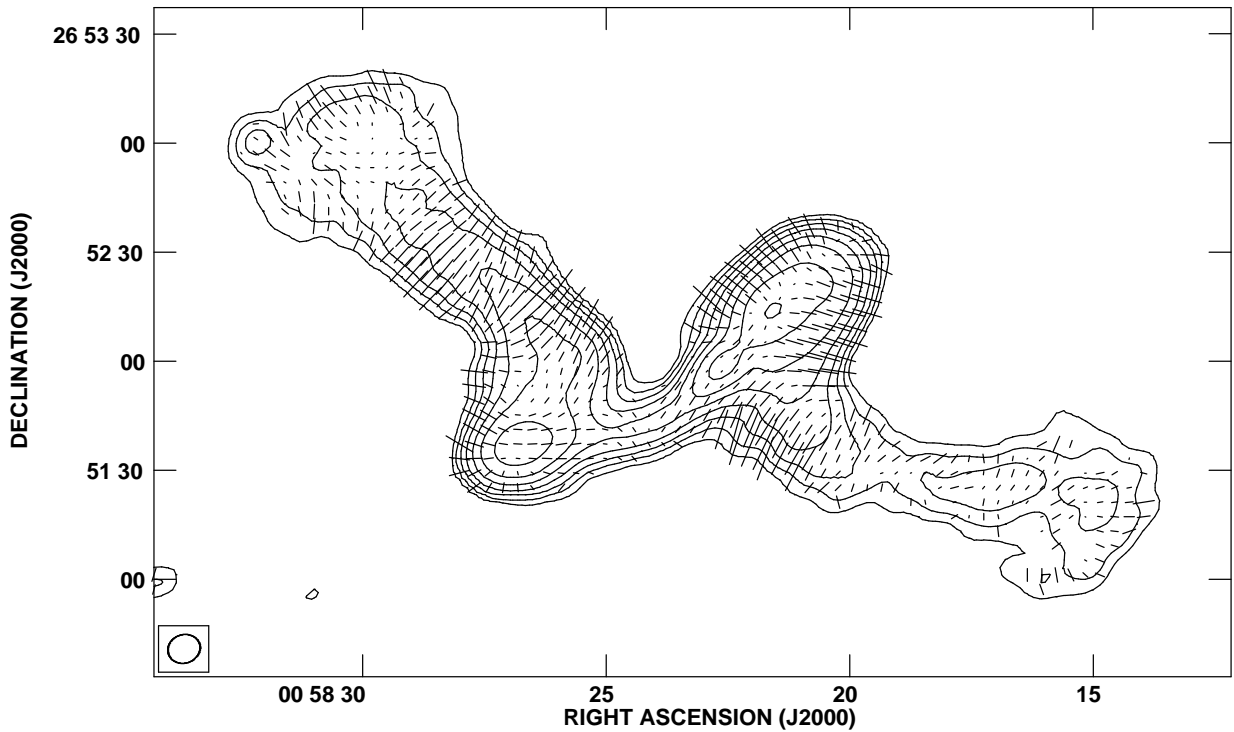}
\caption[]{8.5 GHz D array image of NGC326. The contour levels are
 0.15, 0.34, 0.69, 1.4, 2.9, 5.8, 12, 24 mJy (beam area)$^{-1}$ and the
 restoring beam is $9\arcsec \times 7\farcs 8$. The vectors lengths are 
proportional to the degree of polarization, with 100 per cent corresponding
 to 20 arcsec on the sky, and their directions are those of the E-vector. The 
 unresolved source at the tip of the east wing is a flat-spectrum background
 object.}
\label{DPPERC}
\end{figure*}

\begin{figure}
\begin{center}
\includegraphics[width=8.5cm, angle=0]{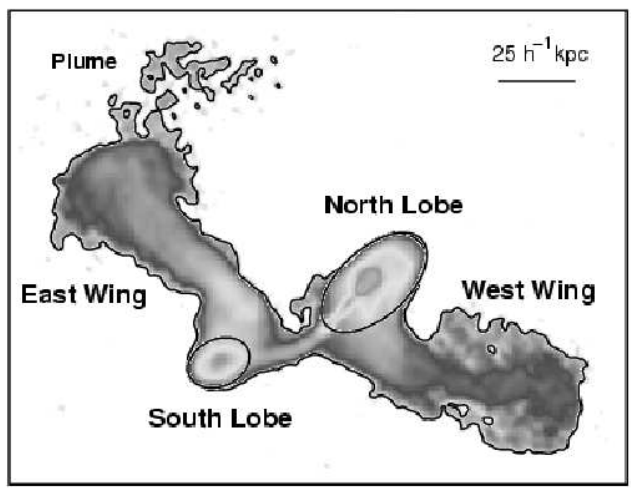}
\end{center}
\caption[]{Source regions referred to in the text.}
\label{notation}
\end{figure}

\begin{figure*}[t]
\includegraphics[width=18cm]{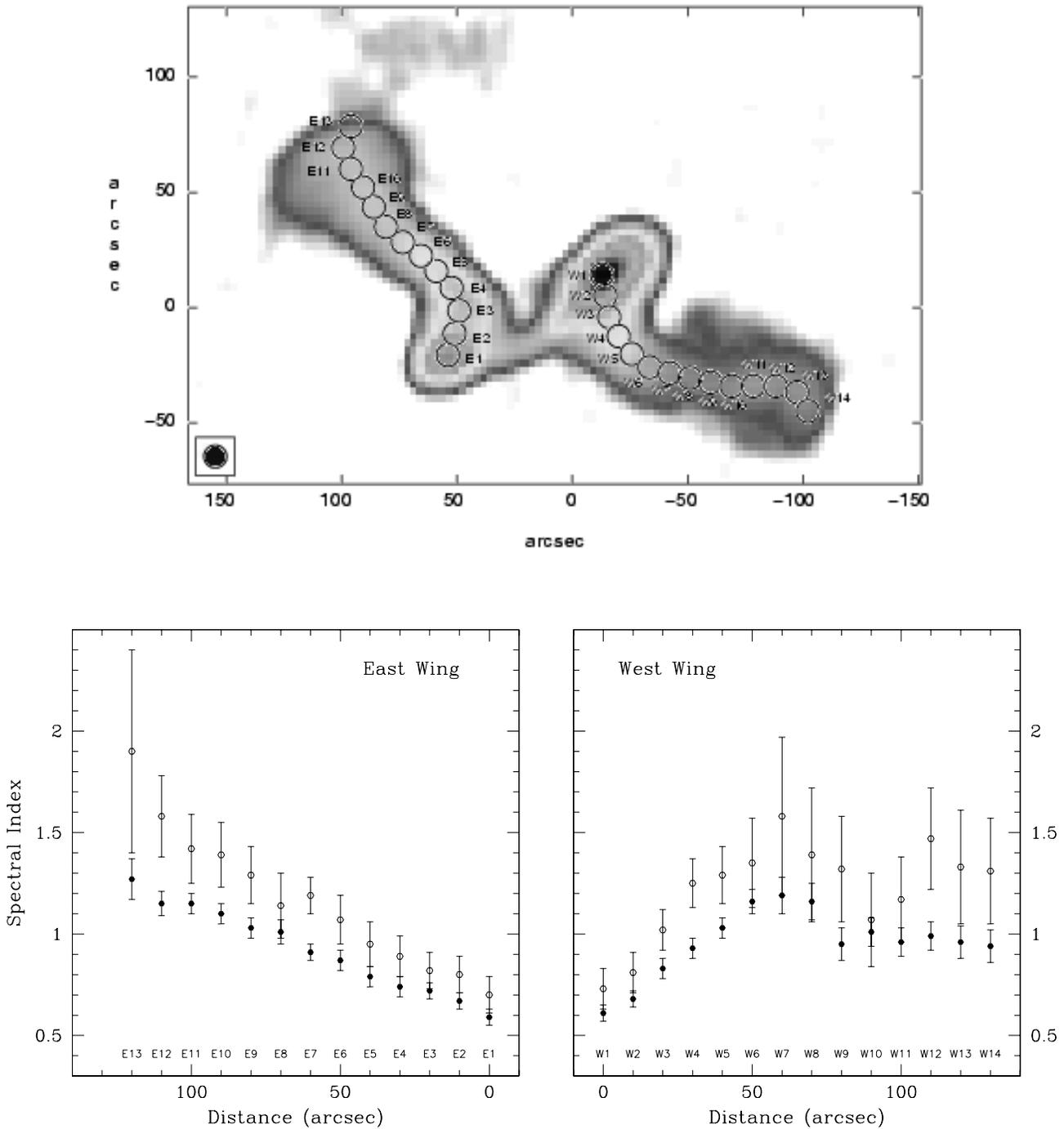}
\caption[]{Spectral index profiles along the wings ridge lines.
 The filled and open dots refer to $\alpha^{1.4}_{4.8}$ and
 $\alpha^{4.8}_{8.5}$, respectively.}
\label{SPXWINGS}
\end{figure*}

\begin{figure*}[t]
\begin{center}
\includegraphics[width=13.5cm, angle=-90]{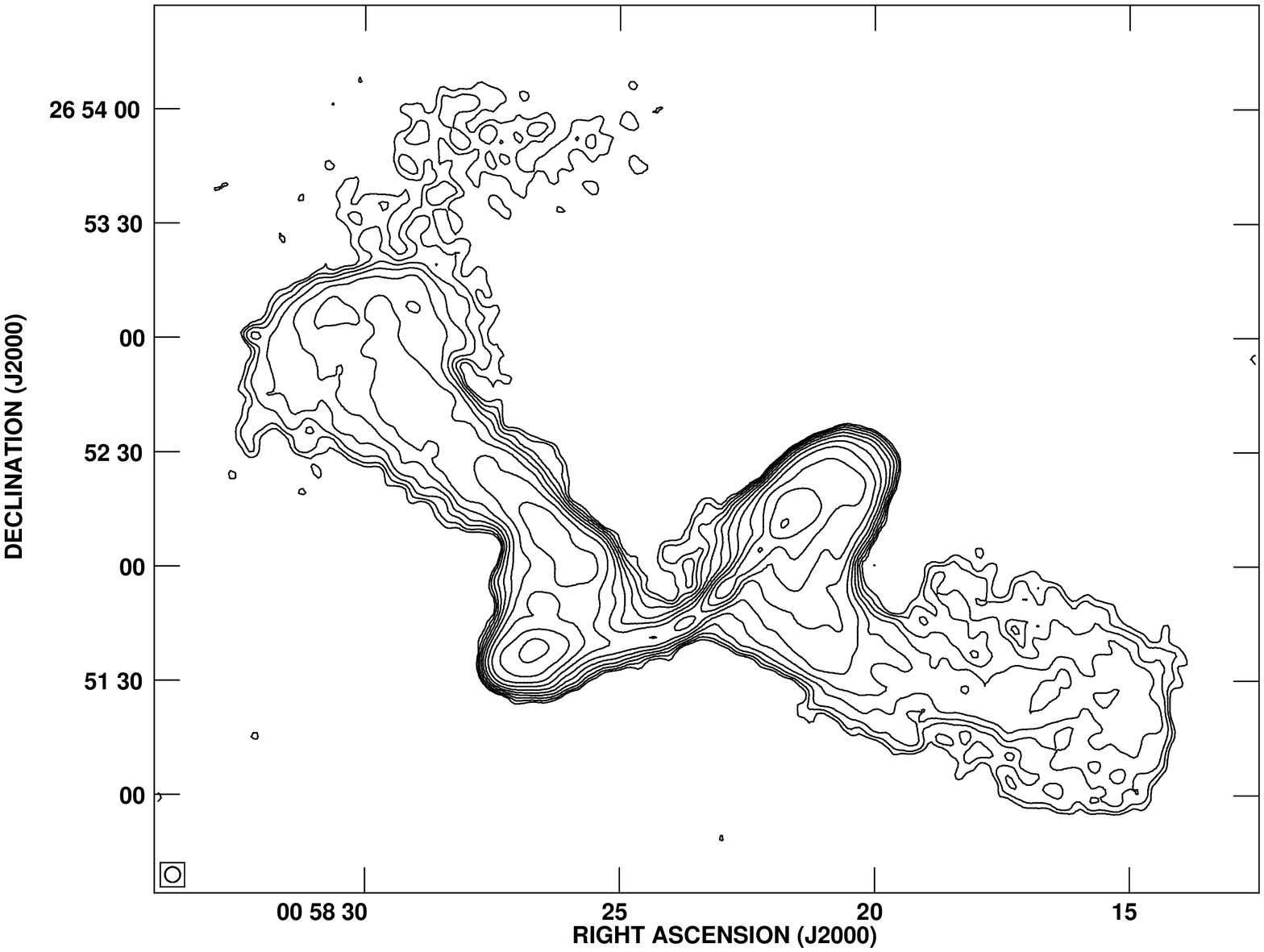}
\end{center}
\caption[]{1.4 GHz A+C array. The contour levels are
0.25, 0.39, 0.61, 0.96, 1.5, 2.3, 3.7, 5.7, 9, 14, 22 mJy 
 (beam area)$^{-1}$ and the restoring beam is $4\arcsec \times 4\arcsec$.}
\label{AC.14.4x4}
\end{figure*}

\subsection{Wings and plume}
The lower resolution images give us information on the extended structure
of the source. At a resolution of about 10\arcsec ~the lobes are 
slightly resolved and the most prominent components are the two wings,
 see Figs.~1 and 2. 

The wings bend and extend away from the lobe axis almost by 
the same extent ($2^{\prime}$).
As already noted by Worrall et al. (1995),
the overall `Z'-shape symmetry of the source is broken by a low surface 
brightness plume located just above the end of east wing. This plume, 
evident only in the 20~cm and 6~cm images, does not follow the source 
symmetry. Fig.~1 shows a galaxy which is located, in projection, in the
 middle of the plume. This corresponds to galaxy G6 of Werner et al. (1999)
 with a radial velocity difference of 1026 km~s$^{-1}$ with respect to the 
center of the cluster. If G6 is a cluster member it moves very fast 
relatively to the intra-cluster medium (slightly more than twice the 
line-of-sight velocity dispersion). Since the sound speed in the cluster is 
$\sim 650$ km~s$^{-1}$, the motion of G6 would be supersonic.
 Anyway, the association of G6 to the plume is unclear. 

Fig.~2 presents the 8.5 GHz map, with the E-field vectors superimposed 
(not corrected for RM, see below). The mean polarization percentage
 (10 \arcsec beam) in the wings is $\simeq4$\% at 1.4 GHz and
 $\simeq 17$\% at 4.8 and 8.5 GHz.
 Both wings show depolarization between 4.8 and 1.4~GHz: the mean values
 of $DP^{1.4}_{4.8}$ (10\arcsec beam) are 0.3 and 0.2 for the east and
 west wing respectively. We calculated the RM at 10\arcsec~resolution 
between the frequencies 1.4, 1.6, 4.8 and 8.5~GHz. We found a mean value
 of $-25 \pm 5$ rad~m$^{-2}$, which is consistent with the galactic value
 (\cite{Simard81}), and a standard deviation $\sigma_{\rm RM}=52$
 rad~m$^{-2}$.
 Given this value for the rotation measure, 
the apparent magnetic field  direction at 8.5 and 4.8~GHz is within 
few degrees from the real one.

The apparent magnetic field is longitudinal and highly
 aligned along the main ridge of emission in both the east and, at least in
 the first part of, the west wing. At the wing edges the magnetic
 field bends to a circumferential configuration.

The surface brightness of the plume is too low to give a reliable 
polarization measure.

At a resolution of 10\arcsec, the radio spectrum of the wings between
 1.4 and 8.5~GHz can be computed out to their full extent.
Fig.~4 shows a plots of the two-frequency spectral indices,  
$\alpha^{1.4}_{4.8}$ and $\alpha^{4.8}_{8.5}$, as a function of the
 position along the two wings. The spectral indices have been sampled 
in circular boxes, of the same size as the beam, centered on the
 ridge lines of maximum emission. In the east wing there is a clear
 monotonically steepening of the radio spectrum from the south lobe to the 
end of the wing: $\alpha^{1.4}_{4.8}$ and $\alpha^{4.8}_{8.5}$
 increase from 0.6 and 0.7 up to 1.3 and 1.9, respectively.
The west wing shows a different spectral behavior:
 $\alpha^{1.4}_{4.8}$ and $\alpha^{4.8}_{8.5}$, increase respectively from 
0.6 and 0.7, close to north lobe, up to 1.3 and 1.5, at a distance of
 $\sim 60\arcsec$ (38 $h^{-1}$ kpc). In the remaining part of the wing, 
$\alpha^{1.4}_{4.8}$ and $\alpha^{4.8}_{8.5}$ decrease and saturate
 to 1.0 and 1.3, respectively.

The resolution of our images allows us to trace spectral index trends also
 along directions  perpendicular to the wing ridge lines. There are
 significant lateral spectral index gradients with marked differences
 between the two wings.

The detailed analysis of these spectral index profiles will be presented
 in paper II.

\begin{figure*}
%%BoundingBox: 50 120 560 805
% BLC 238 378; TRC 669 661
\includegraphics[width=17cm, angle=0]{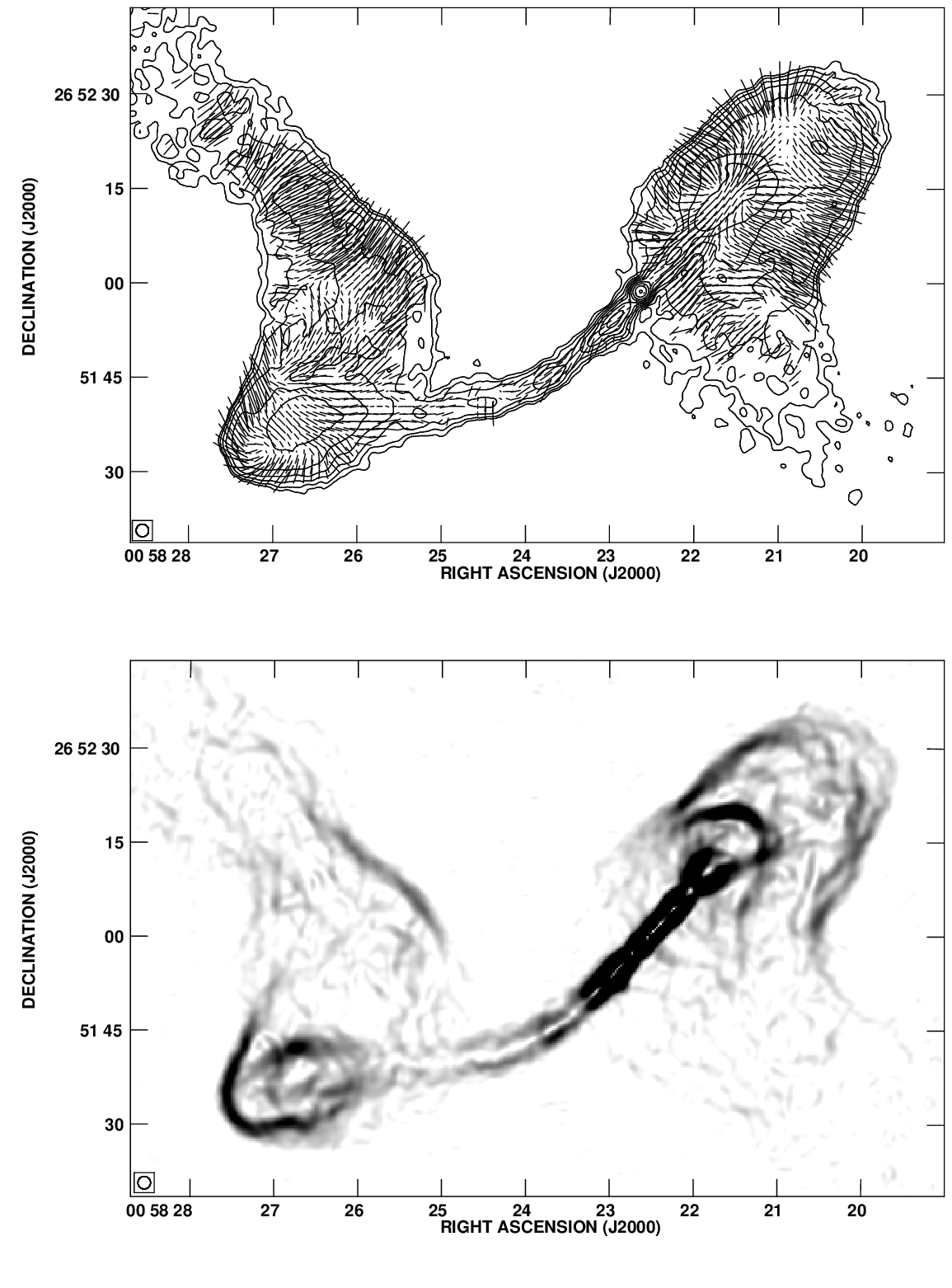}
\caption[]{Top panel: 4.8 GHz B+C array total intensity image.
 The contour levels are 0.14, 0.22, 0.36, 0.58, 0.93, 1.5, 2.4, 3.9, 6.2, 10
 mJy (beam area)$^{-1}$. The vectors lengths are 
proportional to the degree of polarization, with 100 per cent corresponding
to 10 arcsec on the sky, and their directions are those of the E-vector.
Bottom panel: 1.4 GHz A+C array gradient image (AIPS task {\it NINER}). 
In both images the restoring beam is $2\arcsec \times 2\arcsec$. }
\label{BC.48.2x2}
\end{figure*}

\subsection{Lobes}
The images at an angular resolution of 4\arcsec ~can be used to study 
 the shape of the two lobes (see Fig.~5). 
The lobes are quite asymmetric in total emission, extent and
 distance from the core (see also Fig.~6). The southern lobe has an 
ellipsoidal shape, while the northern one is more elongated and wider.
 The north lobe exhibits a shell-like feature which
 is particularly evident in the gradient image shown Fig.~6. This 
  image,  obtained with the AIPS task {\it NINER}, also reveals  a sharp
 frontal border in the southern lobe and a 
straight edge in the east wing. By contrast the northern lobe is
 characterized by well defined lateral edges but has a smooth frontal border.
 Both lobes lack  hotspots. The only structure with a
  significant brightness contrast with respect to the underlying lobe
 is the shell region in the north lobe.

The mean polarization percentage at 8.5 GHz (10\arcsec~beam) is 
12\% and 14\%  in the south and north lobe, respectively (see Fig.~2).
Fig.~6 presents the 4.8~GHz image at 2\arcsec~resolution, with the E-field
 vectors superimposed (not corrected for RM, see Sect.~3.1).
The mean scalar fractional polarization at 4.8~GHz is 23\% and 29\% for 
the south and north lobe, respectively. The variation of the fractional 
point to point
polarization at 4.8~GHz  is strongly 
correlated with the enhancements seen in the total intensity gradient image.
 In the lateral edges of the northern lobe and in the
 frontal head of the southern lobe, the fractional polarization $P_{4.8}$ is
 about 43\%. In the sharp western edge of the east wing $P_{4.8}$ reaches
 a value of $\sim 50$\%. The mean values of $DP^{1.4}_{4.8}$
 (10\arcsec beam) are 0.23 and 0.17 in south and north lobe, respectively, 
while at 2\arcsec~resolution they go up to 0.52 and 0.48.
 In both lobes, the mean RM at 10\arcsec~resolution  is about -20 rad~m$^{-2}$.
 As a consequence the Faraday rotation is negligible at 4.8~GHz and the
 apparent magnetic field in the lobes is circumferential,
 consistently with the configuration deduced from the lower resolution
 8.5~GHz map. The magnetic field configuration is also circumferential 
around the shell structure, where the fractional polarization
 $P_{4.8}$ is about 30\%. 

The mean fractional polarization polarization of the lobes as function of 
frequency and beam area is shown in Fig.~7.
\begin{figure}[b]
%  18 177 564 624
\includegraphics[width=8cm]{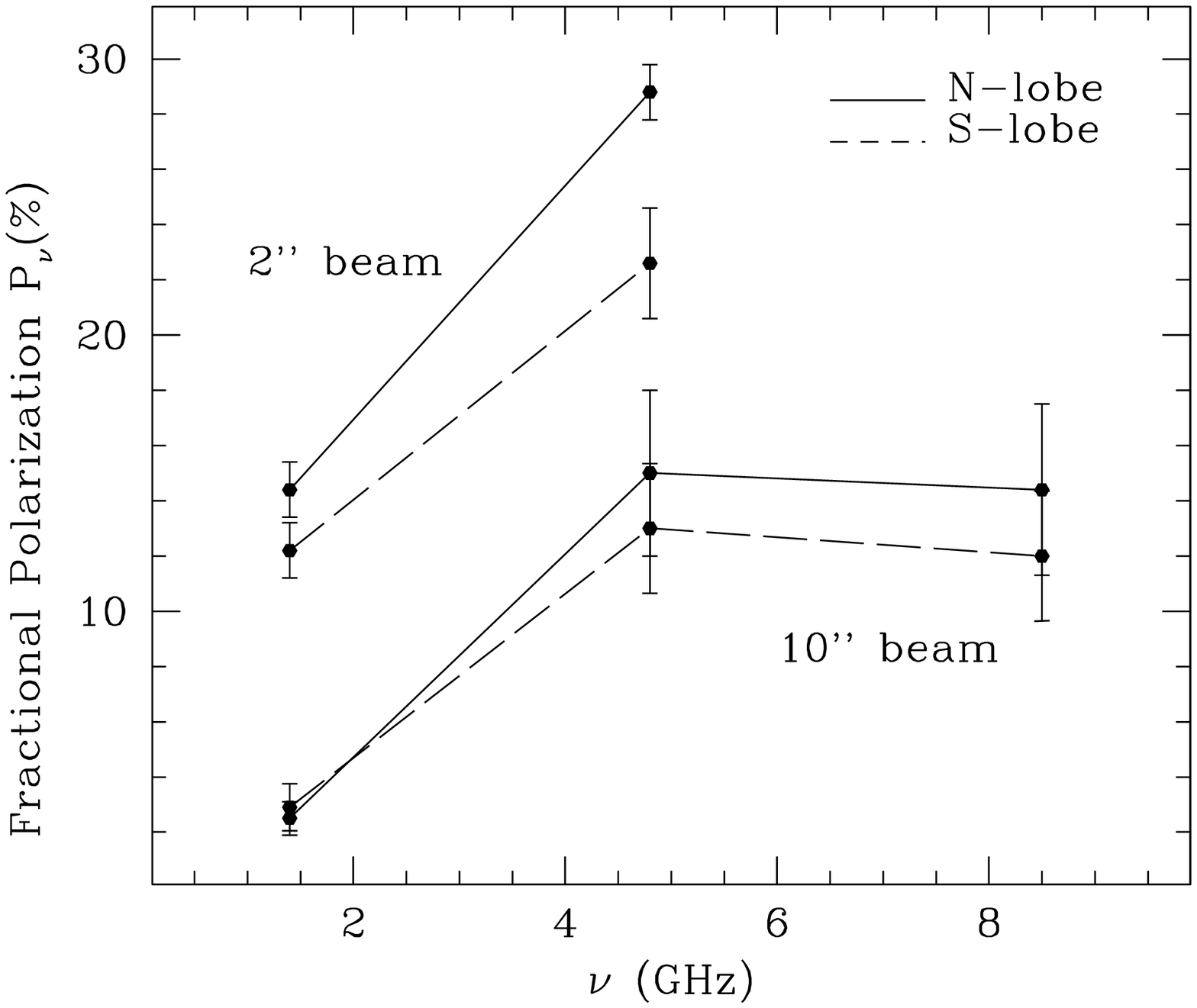}
\caption[]{Mean fractional polarization in the lobes at different 
angular resolution and frequency. The solid and dashed lines refer to
 the north and south lobe, respectively. 
}
\label{lobesfpol}
\end{figure}

\begin{table*}[t]
\caption[]{Lobes and wings properties. The corresponding regions are those
 indicated in Fig.~3.}
\label{lobesandwingstab}
\begin{tabular}{lcccccc}
\hline
\noalign{\smallskip}
& South Lobe   & North Lobe & East Wing & West Wing & Plume\\
\noalign{\smallskip}
\hline
\noalign{\smallskip}
$S_{\rm 20cm}$ (mJy)                  &  213          & 576          & 536            & 288           & 18 \\  
$\theta_{1}\times\theta_{2}$ (arcsec) & $30\times 22$ &$60\times 32$ & $100\times 50$ & $110\times 60$ & $90\times 60$\\
$d_{1} \times d_{2}$ (kpc)            & $22\times 14$ &$38\times 20$ & $63\times 32$ & $ 69\times 38$&$57\times 38$&\\
$\langle\alpha_{4.8}^{1.4}\rangle$  & $0.60 \pm 0.04$&$0.64 \pm 0.02$ &$0.90\pm 0.01$& $1.04\pm 0.01$&$2.37\pm 0.02$\\

$\langle P_{1.4} \rangle$, 10\arcsec beam   &  3\% & 2.5\%  & 6\% &  3\% &$-$\\
$\langle P_{4.8} \rangle$, 10\arcsec beam   &  13\% & 15\%  & 22\%&  16\% &$-$\\
$\langle P_{8.5} \rangle$, 10\arcsec beam   &  12\% & 14\%& 17\%&    14\%&$-$\\

$\langle P_{1.4} \rangle$, 2\arcsec beam   &  12\% & 14\%  & $-$ &  $-$ &$-$\\
$\langle P_{4.8} \rangle$, 2\arcsec beam   &  23\% & 29\%  & $-$ &  $-$ &$-$\\

$\langle DP^{1.4}_{4.8} \rangle$, 10\arcsec beam& 0.23  &  0.17 & 0.27   & 0.18 &$-$\\
$\langle DP^{1.4}_{4.8} \rangle$, 2\arcsec beam & 0.52  &  0.48 & $-$  & $-$ &$-$\\
$\langle RM \rangle$ (rad~m$^{-2}$)&$-21\pm 6$&$-18 \pm 4$ &$ -34 \pm 2$& $-32\pm 6$&$-$\\
$\langle \sigma_{\rm RM} \rangle$ (rad~m$^{-2}$)& 57 & 52  & 35 & 65 & $-$\\
\noalign{\smallskip}
\hline
\end{tabular}
\end{table*}

\begin{figure*}
\includegraphics[width=18cm]{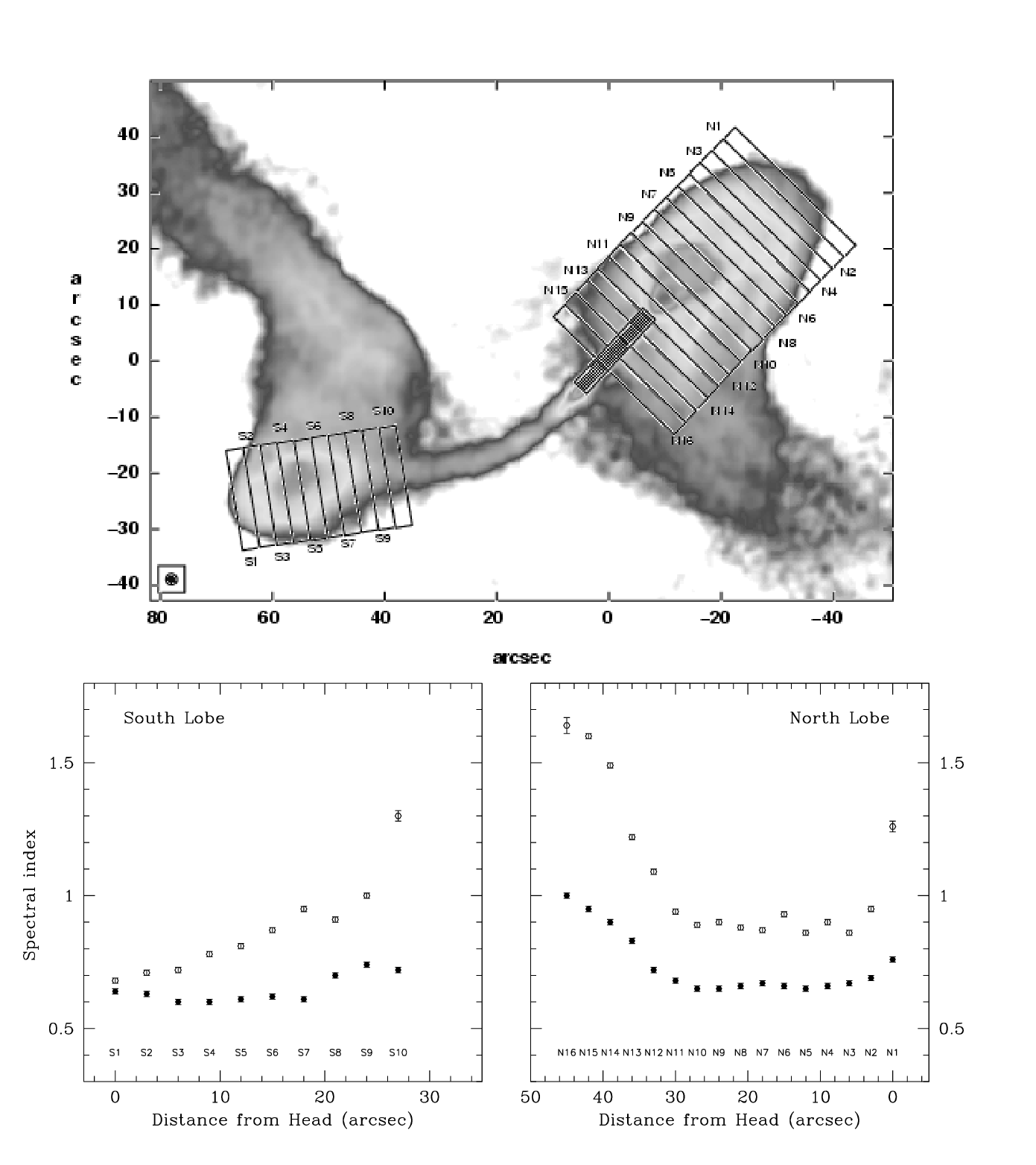}	
\caption[]{Spectral index profiles along the lobes at 2\farcs~ resolution.
 The filled and open dots refer to $\alpha^{1.4}_{4.8}$ and
 $\alpha^{4.8}_{14.9}$, respectively. The shaded box indicates the region
 excluded from the spectral index calculation.}
\label{SPXLOBES}
\end{figure*}

We have computed the average  two-frequency spectral indices
 $\alpha^{1.4}_{4.8}$ and $\alpha^{4.8}_{14.9}$
along slices perpendicular to the lobe axis from the lobe head back to the
 core   using the 2\arcsec~resolution images. 
Slices are 3\arcsec~wide, so that the spectral index measures are 
practically independent. The regions containing the core and the jet in the 
northern lobe have been excluded from the statistics. Fig.~8 shows the
 spectral index trends in the lobes. In the south lobe $\alpha^{1.4}_{4.8}$
 is roughly constant around a value of 0.6 with a moderate
 increase to 0.7 at the lobe end;  $\alpha^{4.8}_{14.9}$
 increases from 0.7  to  1.3. The spectral behavior of the north lobe is 
somewhat complex: $\alpha^{1.4}_{4.8}$ and $\alpha^{4.8}_{14.9}$ decrease
 respectively  from 0.75 and 1.35, at the lobe head, to 0.65 and 0.9, stay
 constant from 10\arcsec~to 30\arcsec, and then increase again up to 1.0 
and 1.6 in proximity of the core. The interpretation of the spectral
 profiles in the north lobe should be considered with caution  since
 it is quite
 possible that the lobe, the jets and part of the west wing are overlapping
 each other because of projection effects.        

The main properties of radio lobes and wings are 
summarized in Tab.~\ref{lobesandwingstab}.

\subsection{Jets}
Images of the jets at a resolution of 2\arcsec ~and 1\farcs2 are shown in 
Fig.~6 and Fig.~9, respectively. The east jet can be
 traced out to a projected distance of 28.4$h^{-1}$ kpc until the south lobe,
 whereas the west jet extends out to 7.6$h^{-1}$ kpc from the core, where
 it suddenly widens and merges into the lobe. Initially both jets are
 straight with a position angle ${\rm p.a.} \simeq 43\degr$. The
 east jet bends gradually (with a curvature radius of about 70\arcsec) and 
approaches the south lobe almost horizontally.\\

The fractional polarization $P_{4.8}$ is about 14\% in the east jet.
 The overlap with the lobe and the wing precludes estimation of the 
fractional polarization in the west jet. The apparent magnetic field 
is transverse in the east jet, whereas in the west jet it is initially
 transverse and becomes longitudinal in correspondence with the shell.\\  
 
We derived the spectral index between 1.4 and 4.8~GHz along the two jets
 as a function of the distance from the core using the  
2\arcsec~resolution images. The resulting trends are shown in Fig.~10. 
The spectral index stays almost constant at a value 
of 0.6 along both jets, although a significant scatter is present in the east 
jet. 

\subsection{Radio cores}
The VLA images at maximum resolution combined with the optical Hubble Space
 Telescope (HST) image of NGC326 (Capetti et al. 2000) show that also the 
secondary nucleus of the dumbbell hosts a radio core (see Fig.~9).
 According to the convention adopted by Werner et al. (1999) for the
 optical cores, hereafter we refer to the two radio cores as `core 1'
 (the radio galaxy) and `core 2' (the companion galaxy). We measure an
 angular distance between the two radio cores of 7\farcs1 which
 corresponds to a projected separation of 4.8 $h^{-1}$ kpc. The luminosity
 of core 1 is typical for a radio galaxy of this total power at 1.4~GHz
 (\cite{Feretti84}). Core 2 is an order of magnitude fainter and, at the
 sensitivity limits of our observations, does not posses radio jets on
 kpc scale. 

\begin{figure*}[t]
% 48 117 532 586 14.5cm
\begin{center}
\includegraphics[width=13.5cm, angle=0]{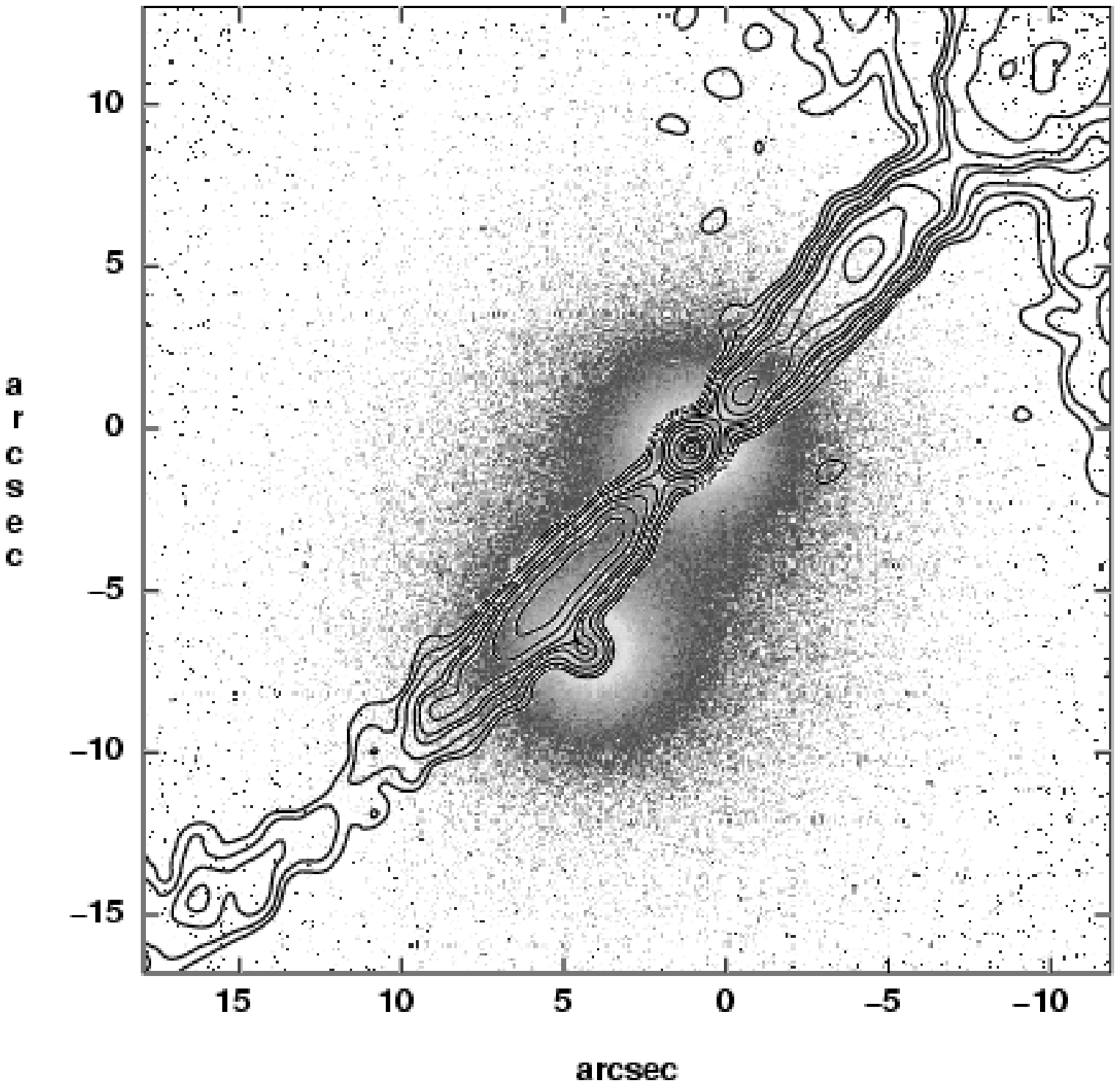}
\end{center}
\caption[]{1.4 GHz A array (contours) overlaid with the HST image 
 of the dumbbell galaxies (grayscale). The contour levels are
0.23, 0.33, 0.48, 0.71, 0.85, 1.1, 1.6, 2.3, 3.4, 4.8
 mJy (beam area)$^{-1}$ and the restoring beam is $1\farcs2 \times 1\farcs2$.}
\label{VLA+HST}
\end{figure*}

\begin{figure}[h]
\includegraphics[width=9cm]{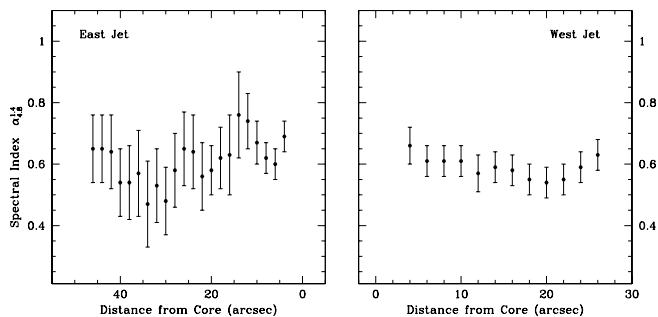}
\caption[]{Profiles of the spectral index between 1.4 and 4.8~GHz along the
 jets.}
\label{JETSSPIX}
\end{figure}
The radio spectra of the two cores between 1.4 and 14.9~GHz are shown 
in Fig.~11. Core 1 has a convex radio spectrum which peaks at about 10~GHz.
 Core 2 has a powerlaw radio spectrum  with an index of 0.3.

Table ~6 summarizes the parameters of the two cores derived from the  images 
at 1\farcs2 resolution.

\section{Collimation and surface brightness of the jets}
We analyzed the evolution of jet collimation and surface brightness as a
 function of the distance from core.
\begin{figure}[t]
\includegraphics[width=9.25cm]{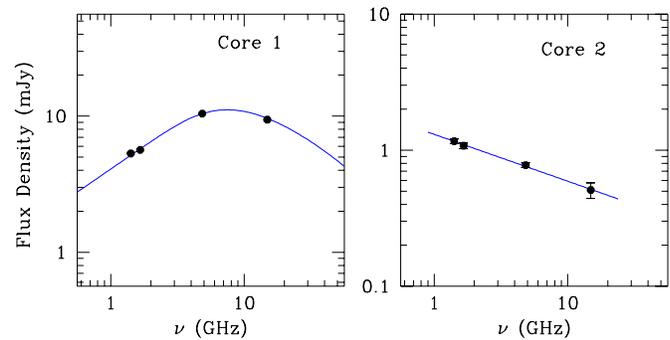}
\caption[]{Radio spectra of the two cores. Core 1 is the radio galaxy.}
\label{CORESPEC}
\end{figure}
\begin{table}[t]
\caption[]{Core properties.}
\label{twocores}

\begin{tabular}{lll}
\hline
\noalign{\smallskip}
&core 1 & core 2\\
\hline
\noalign{\smallskip}
1.4~GHz  position ($\alpha_{\rm J2000})$  &  00$^{h}$58$^{m}$22\farcs6 &  00$^{h}$58$^{m}$22\farcs9\\
1.4~GHz  position ($\delta_{\rm J2000})$ & +26\degr51\arcmin58\farcs7 & +26\degr51\arcmin52\farcs7\\
$S_{\rm {1.4~GHz}}$~(mJy) 		  &$5.33\pm  0.04$ & $1.16\pm 0.04$\\ 
$S_{\rm {4.8~GHz}}$~(mJy)                 &$10.42\pm 0.03$ & $0.78\pm 0.03$\\
$\alpha_{4.8}^{1.4}$                      & -0.5            &0.3\\
$P_{\rm {1.4~GHz}}$~($h^{-2}$WHz$^{-1}$)  & $1.2\times10^{22}$&$2.7\times10^{21}$ \\ 
$P_{\rm {4.8~GHz}}$~($h^{-2}$WHz$^{-1}$)  & $2.3\times10^{22}$&$1.8\times10^{21}$ \\ 
\noalign{\smallskip}
\hline
\end{tabular}
\end{table}

In order to quantify the variations of surface brightness and width along 
the jets, we produced  intensity profiles perpendicular to the jet 
axis for distances up to 40\arcsec~and 20\arcsec~from the core, for the
 east and west jet, respectively.
After having removed the baseline, the 
transverse brightness profiles are well fitted by a simple Gaussian over 
most of the jet length.

 We used both the 2\arcsec~and 1\farcs2 resolution images at 20~cm.
We measured the peak surface brightness $I_{0}$ and FWHM $\Phi_{0}$ and
 derived the deconvolved values $I$ and $\Phi$ by the first-order corrections:
\begin{eqnarray*}
&&\Phi=(\Phi_{0}^{2}-\Phi_{b}^{2})^{1/2}  \\ 
&&I=I_{0}\cdot(\Phi_{b}^{2}/\Phi^{2}+1)^{1/2}   
\end{eqnarray*}
(Killeen, Bicknell \& Ekers 1986) where  $\Phi_{b}$ is the FWHM of the
 restoring beam. Fig.~12 displays the variation of $\Phi$  and
 $I$ as function of distance $\Theta$ along the jets.
 The steps of the transverse cuts are 1\arcsec~and 0.6\arcsec~in the 
2\arcsec~and 1\farcs2 resolution images, respectively. 
The error bars represent the formal 1-$\sigma$ confidence interval given by
 the fitting procedure and do not include the uncertainty introduced by the
 baseline subtraction.  

\begin{figure*}[t]
\includegraphics[width=18cm]{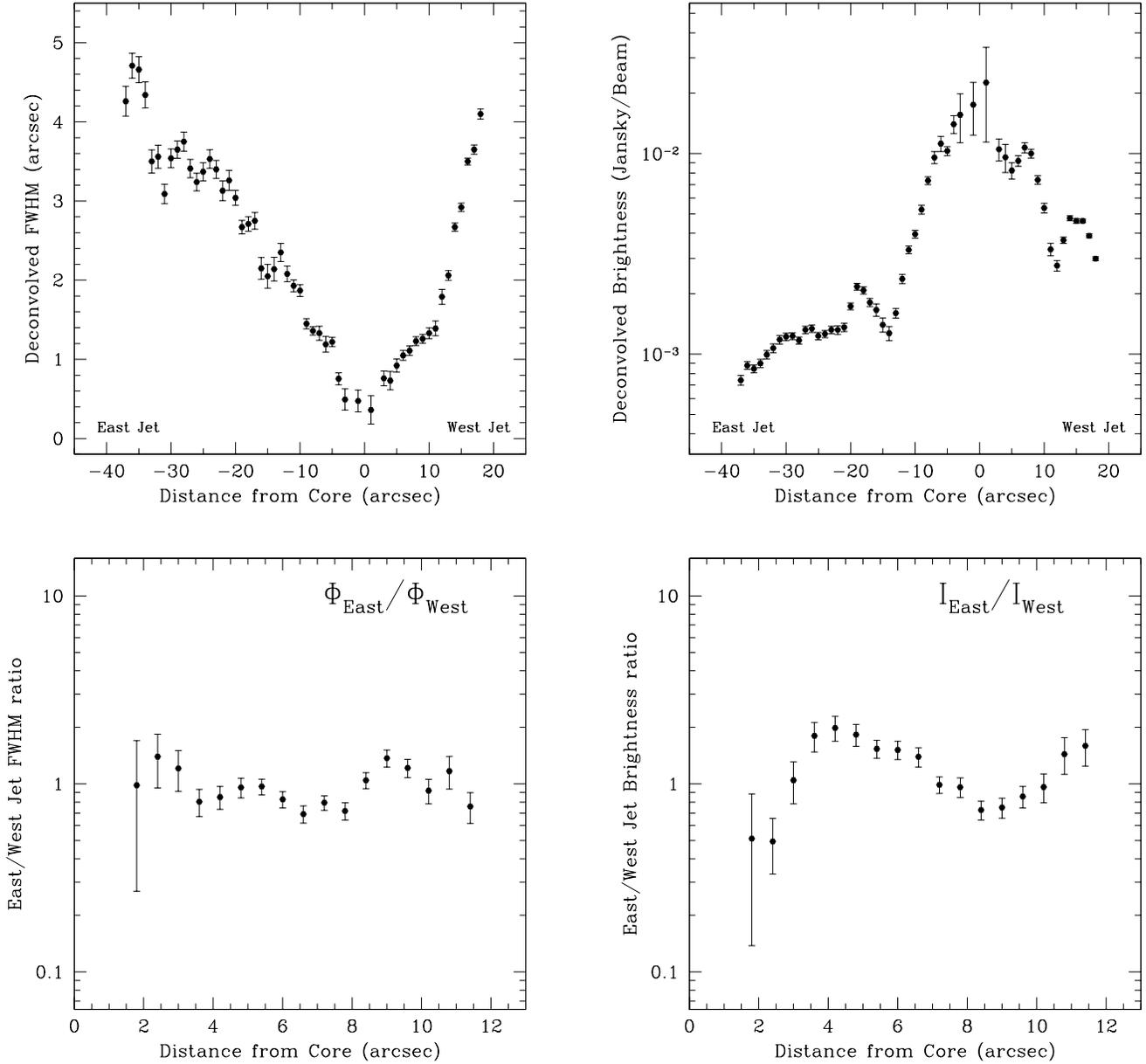}
\caption[]{Top: deconvolved jets FWHM $\Phi$ (left panel) and brightness 
$I$ (right panel) profiles from the 2\arcsec~resolution 20~cm image.
 Bottom: deconvolved FWHM (left panel) and brightness (right panel) ratios
 from the 1\farcs2 resolution 20~cm image.}
\label{JETSPROFILES}
\end{figure*}

Apart for the initial region ($\Theta<$12\arcsec), where the jets are quite
 symmetric (see below), the dependence of $\Phi$ and  $I$ on $\Theta$
 are very different for the east and the west jet.
 The east jet spreads gradually up to 40\arcsec~from the core, but has local
 regions that deviate significantly from the mean. Its deconvolved surface
 brightness  drops by an order of magnitude in going  from the core
 to a distance of 15\arcsec. Then, after a local knot of emission at
 $\Theta=$20\arcsec, $I$  decreases slowly for the remaining 20\arcsec.
 On the contrary, the west jet expands abruptly at 12\arcsec~from the core. 
Its deconvolved brightness initially decreases to a local minimum at
 $\Theta=$12\arcsec~and then raises again peaking at $\Theta=$14\arcsec.
 Beyond this peak the jet decollimates originating the shell-like structure.

The east/west width ratio within 12\arcsec~from the core has 
a mean value of 0.98 and a range from  0.7 to 1.4. The surface brightness
 ratio has a mean value of 1.2 and a range from 0.49, at 2\farcs4 from the
 core, to 2, at 4\farcs2 from the core. The overall jet symmetry 
for $\Theta<$12\arcsec~suggests that the jet axes, at least at their base,
 lie in the plane of the sky. The asymmetry observed at larger distances
 may be due to strong projection effects and/or  differences in the
 density of the gas surrounding the source.\\

Both jets exhibit an average flat $I(\Phi)$ relation 
($I \propto \Phi^{-1.35}$, see Fig.~13).
 This behaviour indicates either that they depart significantly from the 
 assumption of adiabatic flow, or they must be subject to a rapid
 longitudinal deceleration (Fanti et al. 1982).  For example, for a jet with
 no longitudinal component of magnetic field and a spectral index of 0.6 the 
adiabatic condition implies $I \propto \Phi^{-3.4} v_{\rm j}^{-3}$,
 where $v_{\rm j}$ is the jet velocity; in this case the required
 deceleration would be $v_{\rm j}\propto \Phi^{-0.68}$.

\section{Physical parameters of the source}
Using the ``minimum energy assumption'', we calculated with standard formulae
 (\cite{Pacho70}) the equipartition internal energy ($U_{\rm tot}$), the 
energy density ($u_{\rm min}$) and the magnetic field ($B_{\rm eq}$) 
for the jets and the extended components. We also
 calculated the equipartition pressure $p_{\rm eq}=u_{\rm min}/3$.
We assumed a volume filling factor of unity, equal 
energy in relativistic electrons and protons and a radio spectrum ranging
 from 10~MHz to 100~GHz with a spectral index of 0.6.

\subsection{Lobes and wings}
We assumed an ellipsoidal geometry for the lobes and a cylindrical geometry
 for the wings. Sizes for lobes and wings are those  
 reported in Table ~5. The global minimum energy
 parameters are listed in Table ~7. We further calculated
 the variation of the equipartion parameters in ten regions in the
 wings (see Table ~8 and Fig.~14). Worrall \& Birkinshaw (2000) 
reported, for the cluster surrounding NGC326, a core radius
 of $\sim 171 ~h^{-1}$ kpc and a pressure ranging from $7\times10^{-12}
 ~h^{1/2}$ dyne~cm$^{-3}$ to $5\times10^{-12}~h^{1/2}$ dyne~cm$^{-3}$
 at 100 kpc from the source.

\begin{table}[h]
\label{extendedFields}
\caption[]{Global equipartition parameters.}

\begin{tabular}{lllll}
\hline
\noalign{\smallskip}
  	& $U_{\rm tot}$ &  u$_{\rm min}$ & P$_{\rm eq}$ &  B$_{\rm eq}$  \\
        & erg          & erg/cm$^{3}$  & dyne/cm$^{2}$ &  $\mu$G         \\
\noalign{\smallskip}
\hline
\noalign{\smallskip}
N-Lobe &$6.2\times10^{56}$& $9.3\times10^{-12}$ & $3.1\times10^{-12}$&10.0\\
S-Lobe &$3.4\times10^{56}$& $5.1\times10^{-12}$ & $1.7\times10^{-12}$&7.4\\
W-Wing &$3.3\times10^{57}$& $1.4\times10^{-12}$ & $4.8\times10^{-13}$&3.9\\
E-Wing &$3.1\times10^{57}$& $2.1\times10^{-12}$ & $6.9\times10^{-13}$&4.7\\
\noalign{\smallskip}
\hline
\end{tabular}

\end{table}

\begin{table}[h]
\label{wingsFields}
\caption[]{Wings equipartition parameters ($\alpha=0.6$).}

\begin{tabular}{llll}
\hline
\noalign{\smallskip}
distance &  u$_{\rm min}$ & P$_{\rm eq}$ &  B$_{\rm eq}$\\
arcsec   &  erg/cm$^{3}$  & dyne/cm$^{2}$ &  $\mu$G  \\
\noalign{\smallskip}
\hline
West Wing&&&\\
\noalign{\smallskip}
26 & $2.3\times10^{-12}$ & $7.7\times10^{-13}$&5.0\\
36 & $1.7\times10^{-12}$ & $5.7\times10^{-13}$&4.3\\
43 & $8.2\times10^{-13}$ & $2.7\times10^{-13}$&3.0\\
53 & $7.5\times10^{-13}$ & $2.5\times10^{-13}$&2.8\\
62 & $7.1\times10^{-13}$ & $2.4\times10^{-13}$&2.8\\
72 & $7.3\times10^{-13}$ & $2.4\times10^{-13}$&2.8\\
83 & $8.2\times10^{-13}$ & $2.7\times10^{-13}$&3.0\\
93 & $8.1\times10^{-13}$ & $2.7\times10^{-13}$&2.9\\
103 & $7.7\times10^{-13}$ & $2.6\times10^{-13}$&2.9\\
113 & $1.8\times10^{-12}$ & $5.9\times10^{-13}$&4.4\\
\hline
East Wing&&&\\
16 & $3.3\times10^{-12}$ & $1.1\times10^{-12}$&5.9\\
28 & $3.6\times10^{-12}$ & $1.2\times10^{-12}$&6.2\\
38 & $3.6\times10^{-12}$ & $1.2\times10^{-12}$&6.2\\
45 & $2.2\times10^{-12}$ & $7.5\times10^{-13}$&4.9\\
53 & $2.0\times10^{-12}$ & $6.8\times10^{-13}$&4.7\\
63 & $2.0\times10^{-12}$ & $6.7\times10^{-13}$&4.6\\
73 & $1.6\times10^{-12}$ & $5.4\times10^{-13}$&4.2\\
82 & $1.2\times10^{-12}$ & $3.9\times10^{-13}$&3.6\\
93  & $1.2\times10^{-12}$ & $4.0\times10^{-13}$&3.6\\
103 & $1.4\times10^{-12}$ & $5.2\times10^{-13}$&4.1\\
\noalign{\smallskip}
\hline
\end{tabular}

\end{table}

Therefore the inner radio lobes appear to be close to
the pressure of the external gas while the wings appear to be under-pressured
by a factor of 10.

\subsection{Jets}
We computed the minimum energy magnetic field and pressure in the jets 
using the deconvolved FWHM, $\Phi$, and surface brightness, $I$, as derived
 in Sec.~4. We assumed for each segment of the jet a cylindrical shape
 with a cross section and length of $\pi \Phi^{2}$ arcsec$^{2}$ and 1
 arcsec, respectively. The minimum energy magnetic field varies
 from 40 to 9 $\mu G$ going from the core outwards. The equipartition 
pressure varies from $ 5\times10^{-11}$ to $3\times10^{-12}$ $h^{1/2}$
 dyne/cm$^{2}$ (see Fig.~15).

From the X-ray luminosity reported by Worrall et al. (1995) for the hot
 galactic corona (i.e. $10^{41} ~h^{-2}$ erg/sec), we computed the expected
 external pressure on the jets. We assumed for the corona a temperature
 of 1~keV and a gas density profile $n=n_{0}\cdot\left(1+(r/r_{c})^{2}
\right)^{-b}$, where $n_{0}$ is the central density while
 $r$ and $r_{c}$ are the distance from the galaxy and the corona core radius,
 respectively. We found that the jets are roughly in pressure equilibrium
  with the surrounding gas (see Fig.~15) for a corona core 
radius $r_{c}= 3~h^{-1}$ kpc and $b=3$; in this case the central density 
results $n_{0}\simeq 0.2 ~h^{1/2}$ cm$^{-3}$. A model with
 $r_{c}= 2~h^{-1}$ kpc, $b=2$ and  $n_{0}\simeq 0.3 ~h^{1/2}$ cm$^{-3}$ gives
 an equivalent result.

\begin{figure}[t]
\begin{center}
\includegraphics[width=8cm]{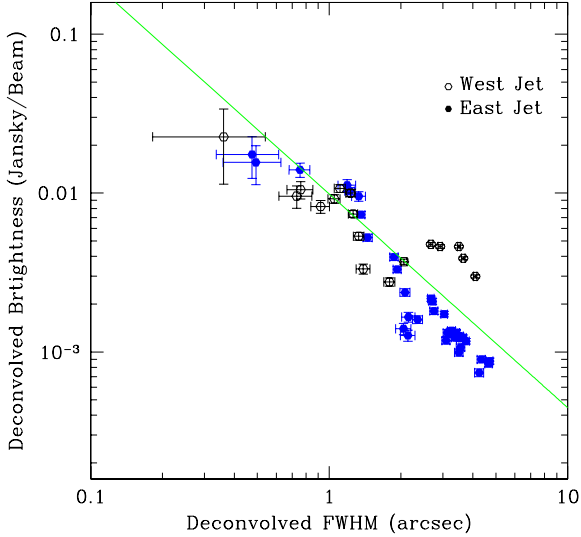}
\end{center}
\caption[]{
Deconvolved brigthness  as a function of the deconvolved FWHM, $I(\Phi)$,
 for the west (open dots) and east (filled dots) jet. The line  
  has a slope of -1.35.}
\label{WP}
\end{figure}

\begin{figure}[b]
\begin{center}
\includegraphics[width=9cm]{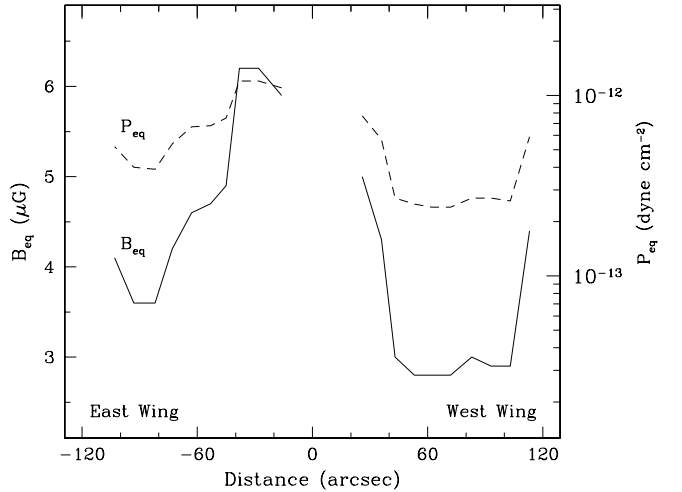}
\end{center}
\caption[]{Minimum energy magnetic field (solid line, left y-axis) 
and pressure (dashed line, right y-axis) in the wings.}
\label{WINGSEQ}
\end{figure}

\begin{figure}[h]
\begin{center}
\includegraphics[width=9cm]{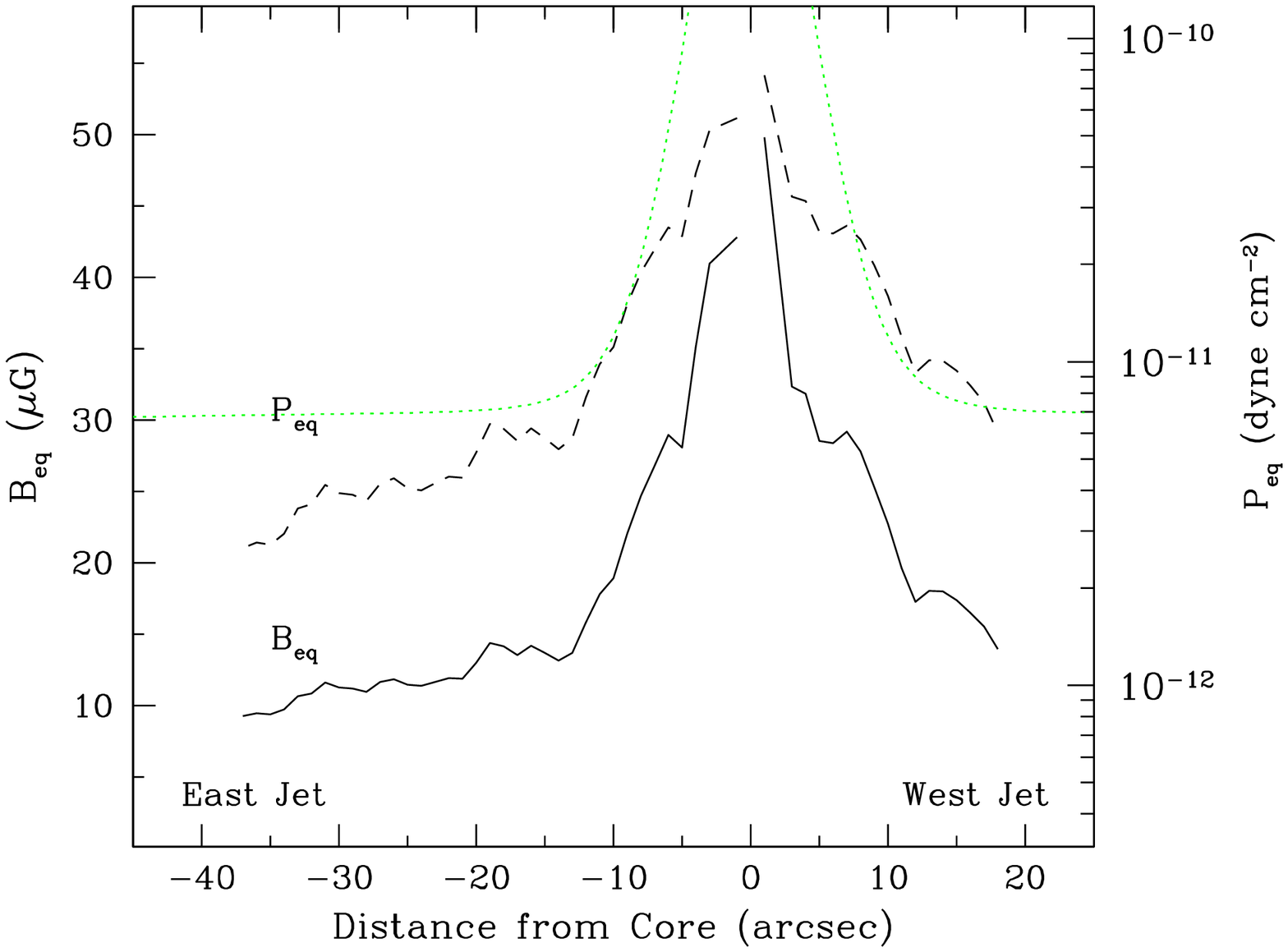}
\end{center}
\caption[]{Minimum energy magnetic field (solid line, left y-axis) 
and pressure (dashed line, right y-axis) in the jets. The dotted line 
represents the expected thermal pressure from the cluster gas and from
 the hot galactic corona model described in the text.}
\label{JETEQ}
\end{figure}

\section{Summary}
We have presented  multi-frequency data of the radio 
galaxy NGC326, together with previously published data that were
 re-analyzed for the purpose of this work. These data allowed us to
 investigate the source morphology at different levels of spatial
 resolutions.

The source structure is complex at all resolutions 
and may be dominated by  projection effects.

At the lowest resolution the twin wings appear fairly symmetric but
 the `Z'-shape is disturbed by a plume of low surface brightness and 
very steep spectral index ($\alpha > 2$). The plume  overlap, in 
projection,  a possible fast-moving member of the galaxy cluster.

At the intermediate resolution of about 4\arcsec~the source lobes appear
 asymmetric in shape, size, brightness and projected distance from the core.
No compact components (hostspots) 
are present in the lobes.

At higher resolution, the jets are very symmetric in both flux density and
 width within the inner 12\arcsec~from the core. This indicates that the
 axis of the jets, at least at their base, lies on the plane of the sky.
 Beyond this distance, the east jet bends gradually with a curvature radius 
of about 70\arcsec~reaching the south lobe. The west jet is straight but,
 after  recollimation expands abruptly, giving rise to a shell-like
 structure embedded in the north lobe.

Overlapping our highest resolution VLA images of the core region with the
 HST image of the dumbbell, we found that also the companion of the radio
 galaxy hosts a radio core.

 NGC326 is significantly polarized at all frequencies and
 resolutions. The magnetic field configuration is circumferential in the 
lobes and highly aligned with the ridge of emission in the wings.\\
 
We traced the profile of the spectral index between 1.4 and 8.5 GHz for
 the entire wing  length. The spectral index increases gradually going 
from the regions in proximity of the lobes to the wing ends. The spectral
 index distribution in the lobes is complicated, presumably due to the mixing
 with the wings. Both jets show a constant spectral index profile with
 a mean value of 0.6.

Finally, we computed the minimum energy, magnetic field and pressure in
 wings, lobes and jets. We found that the lobes are close to equilibrium
 with the external gas pressure while the wings are under-pressured by
 a factor of 10.

These observations allow us to investigate in detail the spectral and
 polarizarion properties of this peculiar object at different levels of
 spatial resolution and constitute an excellent data set which
 permits  testing the ``standard'' theory of the synchrotron aging analysis.

The interpretation of the data will be discussed in papers II and III.

\begin{acknowledgements}
The National Radio Astronomy Observatory is operated by Associated 
Universities, Inc., under contract with National Science Foundation.
\end{acknowledgements}


\begin{thebibliography}{}

\bibitem[Baars et al. (1997)] {Baars} Baars, J. W. M., Gendel, R.,
 Pauliny-Toth, I. I. K., Witzel, A., 1977 A\&A 61, 99

\bibitem[Battistini et al. 1980]{Battistini80} Battistini, P., Bonoli, F.,
 Silvestro, S., Fanti R., Gioia, I. M., and Giovannini, G., 1980, A\&A 85, 101

\bibitem[Blandford \& Icke 1978] {BlandfordIcke78} Blandford, R. D., \& Icke,
 V., 1978, MNRAS, 185, 527

\bibitem[Capetti et al. 2000] {Capetti00} Capetti, A., de Ruiter, H. R.
, Fanti, R., Morganti, R., Parma, P., Ulrich, M.-H., 2000, A\&A 362, 871

\bibitem[Colla et al. 1975]{Colla75} Colla, G., Fanti, C., Fanti, R., Gioia,
 I., Lari, C., Lequeux, J., Lucas, R., Ulrich, M. H., 1975, A\&AS 20, 1 


\bibitem[Davoust \& Consid\`ere 1995]{Davoust95} Davoust, E., \&  Consid\`ere,
 S., 1995, A\&AS, 110, 19

\bibitem[Ekers et al. 1978]{Ekers78} Ekers, R. D., Fanti, R., Lari, C.,
 Parma, P., 1978 Nature, 276, 588

\bibitem[Ekers 1982]{Ekers82} Ekers, R. D., 1982 in {\it Extragalactic
 Radio Sources}, Proceeding of IAU Symposium No. 97, 465

\bibitem[Fanti et al. 1977]{Fanti77} Fanti, C. Fanti, R., Gioia , I. M.,
 Lari, C., Parma, P., Ulrich, M. H., 1977, A\&AS 29, 279

\bibitem[Fanti et al. 1982]{Fanti82} Fanti, R., Lari, C., Parma P., Bridle,
 A. H., Ekers, R. D., Fomalont, E. B., 1982, A\&A 100, 169  

\bibitem[Feretti et al.  1984] {Feretti84} Feretti L., Giovannini G.,
 Gregorini L., Parma P., Zamorani G., 1984, A\&A, 139, 55 

\bibitem[Fomalont 1981] {Fomalont81} Fomalont, E. B., 1981 in {\it Origin of
 Cosmic Rays} ed. Setti, G.,  Spada, G., and Wolfendale, A. W., (Reidel,
 Boston), p. 111 

\bibitem[Killeen et. al 1986] {killeen86} Killeen, N. E. B., Bicknell, G. V.,
 Ekers, R. D., 1986, ApJ, 302, 306 

\bibitem[Murgia 2001] {Murgia 2001} Murgia M., 2001, Ph.D. Thesis, University
 of Bologna  

\bibitem[Pacholczyk 1970] {Pacho70} Pacholczyk A.G., 1970, `Radio
 Astrophysics', Freeman \& Co., San Francisco

\bibitem[Parma et al. 1991] {Parma91} Parma, P., Cameron, R. A., and de
 Ruiter, H. R., 1991, AJ 102, 1960 

\bibitem[Rees 1978]{Rees78} Rees, M. J., 1978, Nature 275, 516 

\bibitem[Sargent (1973)]{Sargent73} Sargent, W. L. W., 1973, Astrophys.J.
 Lett. 182, L13

\bibitem[Simard-Normandin et al. 1981]{Simard81} Simard-Normandin M.,
 Kronberg P. P., Button S., 1981, ApJS, 45, 97S

\bibitem[Wirth et al. (1982)]{Wirth82} Wirth, A., Smarr, L., Gallagher,
 J. S., 1982, AJ 87, 602

\bibitem[Werner et al. 1999]{Werner99} Werner, D. M., Worrall, D. M.,
 and  Birkinshaw, M.,  1999, MNRAS 307, 722

\bibitem[Worrall et al. 1995]{Worrall95} Worrall, D. M., Birkinshaw,
 M., Cameron , R. A., 1995, ApJ 449, 93

\bibitem[Worrall \& Birkinshaw 2000]{Worrall00} Worrall, D. M.,
 \& Birkinshaw, M., 2000, ApJ 530, 719

\end{thebibliography}
\end{document}